%Paper: hep-ph/9302218
%From: "VSTST2::COMELLI"@BNLDAG.AGS.BNL.GOV
%Date: Thu, 4 Feb 1993 9:23:36 -0500 (EST)

\end{quotation}
\end{titlepage}
%%%%% TITLE PAGE
%------------------------------------------------------------------------
 One of the most exciting (and corageous) predictions of a large
 class of supersymmetric models is the existence of (at least) one
 light Higgs scalar [1], that would be certainly discovered at the next
 colliders ( alternatively, the non observation of such particle
would be rather difficult to explain for the previous models).

An extremely important theoretical task becomes therefore the accurate
determination of a rigorous upper bound for this light Higgs mass.
For renormalizable models, this means that a calculation that includes
radiative corrections e.g. at the one loop level would be relevant and welcome.

A well known and particularly illustrative example of the previous statement
is provided by the so called  Minimal low energy Supergravity Models [2].
Here, as a consequence of Supersymmetry, one has the famous tree level bound:
\begin {equation}
M_{H_{1}} \leq M_{z}
\end {equation}
where by $M_{H_{1}}$ we denote, from now on, the mass of the ligthest CP even
Higgs scalar.This bound, as it has recently been stressed in several
publications
 [3], is common to a class of ``Minimal'' SUSY models, that includes SUGRA
models
 where e.g. all the SUSY breaking scalar masses are supposed to be equal
at the GUT scale [4], but also models where such a constraint is not imposed
and where soft breaking terms can be varied independently.

When radiative corrections to the bound of eq.(1) are computed [5], two
important effects are generated. The first one is a substantial increase
 of the upper bound, essentially due
 to the top$\!-\!$stop contribution to the
Coleman Weinberg [6] effective potential, that contains a quartic
top mass dependence.
The most dramatic consequence of this is that , for $M_{t} \simeq 150 \; GeV$,
$M_{\tilde {t}} \simeq 1\; TeV$, the new bound becomes now
$\simeq 120\; GeV$, that
 is out of the reach of LEP2 with integrated luminosity of
$ 500 \; pb^{-1}$ and $\sqrt{s}=190 \; GeV$ [7].
The second remarkable effect is the fact that both the numerical
value of the improved bound and other important phenomenological features
 (like the  possibility that the CP odd "pseudoscalar " becomes
lighter than $H_1$) become
 now different within the previous large class of  "Minimal" models [8] , that
would be of paramount importance in case of future Higgs(es) discovery.

A few years ago it was shown [9] that another class of models exists for which,
at least over an interesting region of the parameter space, a bound for the
lighthest scalar Higgs mass can be simply and elegantly computed.
Such are those models where an extended gauge symmetry generated by the
 exceptional group $E_6$, embedded into a
supersymmetric minimal spectrum of 27-plets
with fermions of the conventional (quarks and leptons) type, is broken down
 by Hosatani type mechanisms [10] to either a rank 6 or a rank 5 low energy
 residual gauge group (for an exhaustive discussion of the various theorical
 details we defer to the existing literature [11]).

In both cases, Haber and Sher [9] derived bounds at tree level for the lightest
 scalar that are only slightly higher than the corresponding bound for Minimal
 SUSY models eq.(1). In particular, for the simplest rank 5 case with a
minimal content of particles (two Higgs doublets and one Higgs singlet in the
third generation acquiring vevs $v_1,v_2,v_3$ ) that is usually called
(minimal)
$\eta$ model, the result
\begin{equation}
  M_{H_1}\leq 108\; GeV
\end{equation}
was found (similar values obtaining for the rank 6 case).

To see how one can possibly obtain a bound like that of eq.(2) and also for
 better understanding the philosophy of our paper, a quick review of its
 derivation and of the role of the parameters of the scalar sector of the
models is requested.  The latter ones are the coupling constant $ \lambda$,
that
 multiplies the
 trilinear term of the original supersymmetric superpotential ($E_6$
invariance forbidding any other possible e.g. bilinear supersymmetric term),
and the parameters of the SUSY breaking sector, i.e. the scalar mass terms
$\mu_1$,
$\mu_2,\mu_3$ and another soft breaking mass term $\lambda A$.

After imposing the three conditions of minimum and replacing the scalar mass
 terms by the related vevs one is left with four free parameters, since it is
 still possible to relate $v_1^2+v_2^2$ to $M_z^2$ \footnote{We identify
 the physical Z,Z' states with the mathematical $ Z_0\; , Z_0'$ gauge
eigenstates.
 Given the existing bound on $ M_{Z'}$, this makes no practical difference} :
\begin{equation}
   M_z^2=\frac{1}{2} (g_L^2+g_y^2)(v_1^2+v_2^2)= \frac{1}{2} g_z^2 v^2
\end{equation}
A possible choice of the four parameters is given, for instance, by the set :
$$
 \frac{ v_2}{v_1} = \tan{\beta} ,\;\; v_3,\;\; \lambda,\;\; \lambda A.
$$

Alternatively, one can use the Z' mass:
\begin{equation}
M_{z'}^2= \frac{1}{18} g_{\eta}^2 (v_1^2+16 v_2^2+25 v_3^2)
\end{equation}
($ g_{\eta}$ is the extra $U(1)$ coupling,
and in practice $g_{\eta}=g_y$ ) and the mass of
 the single (CP odd) pseudoscalar of the model:
\begin{equation}
  M_{ps}^2=\lambda A \{ \; \frac{v_1v_2}{v_3} + \frac{v_1v_3}{v_2} +
\frac{v_2v_3}{v_1} \; \}
\end{equation}
to replace $v_3$ and $\lambda A$ , which is often done
when numerical analyses are
 presented. The neutral CP even sector of the model contains three physical
 states $H_1,H_2,H_3$. To obtain the expression of the physical masses one has
 to diagonalize a 3x3 mass matrix $M^2$ whose six independent elements $\;
 [m^2]_{ij}=a_{ij}=a_{ji} \;$ can be cast in the form:
\begin{equation}
   a_{ij}= \alpha_{ij} v_3+ \beta_{ij} \;\; \mbox{for}\;\; (i,j)\not= (3,3)
\;\;\;
\mbox{and} \;\;
   a_{33}=  \gamma_{33}v_3^2+ \frac{ \lambda A v_1 v_2}{v_3}
\end{equation}
where $(\alpha,\beta)_{ij}$ and $\gamma_{33}$ do
not depend on $v_3$, and their explicit
 expression is given for instance in ref.[9].
As it was pointed out by Drees [12], the determination of a bound
for the lightest
Higgs mass is strongly affected by the value of the ratio $\lambda A/v_3$.
The values $ \lambda A \ll v_3$ correspond to a certain region in the
parameters space, that we shall refer to for simplicity as the
 "Haber-Sher's region", where the bound of eq.(2) can be derived
without enforcing extra assumptions on the non purely gauge couplings of
the model. Conversely in the "Drees region", where $\lambda A$ is not $\ll v_3$
, the derivation of a bound needs extra assumptions on
the $\lambda$ parameter.
Invoking reasonable renormalization group equations (RGE)
arguments (and assuming $M_t\simeq 40$ GeV), Drees was able to fix a
somehow higher value for the bound in this region,
 qualitatively equal to $M_{H_1}\leq 170$ GeV,
and, strictly speaking, this should be considered
as the true bound of the model ( at least,
for the assumed $M_t$ value).

The origin of the difference between the bounds
in the  two regions can be easily
understood if one uses a simplified procedure based on the assumption
$v_3 \gg v_1,v_2$ ( no statements about $\lambda A/v_3$).
The latter choice is phenomenologically motivated by the most recent
bounds on the mass of the extra Z of the model,
that can be derived either via direct CDF limits [13]
or via indirect analyses of LEP1 data [14], [15],
 both leading to the result:
$$
M_{Z'}\geq 300\;\; GeV \; \simeq 0.4\;\; v_3
$$
from wich $ v_3\gg v_1,v_2$ already emerges
(note that future negative searches of the extra Z at CDF
and LEP2 would soon improve the previous bound by a factor 2 [16]).
In this configuration, one can show that
one "light" Higgs exists i.e. one whose mass ,
which does not become of O($v_3$), is given by the following expression:
\begin{equation}
  M_{H_1}^2= [\beta_{11} \cos^2{\beta} + \beta_{22} \sin^2{\beta} +\beta_{12}
\sin{2 \beta} - \frac{[m_{23}^2 \sin{\beta} +m_{13}^2 \cos{\beta}]^2}
{\gamma_{33}v_3^2} ]
\end{equation}
In the Haber-Sher's region ,$v_3 \gg \lambda A $,
 the negative  contribution coming from the
(1,3) and (2,3) non diagonal matrix elements of  eq.(7) does never vanish.
This  brings a (negative) term of O($\lambda^4$) that, when combined
with the positive O($\lambda^2$) contribution of the remaining
matrix elements, produces a maximum becoming, in the limit
$\tan{\beta} \rightarrow \infty$, the bound of eq. (2).

In the Drees region for $\lambda A \simeq O(v_3)$ it is conversely possible
for the previous negative contribution to
vanish.
This leaves a new bound that contains a positive quadratical O($\lambda^2$)
term and has also a $\tan{\beta}$ dependence.

For $\lambda^2 < g_Z^2/2+g_{\eta}^2/3$ the maximum is the same as in
the  Haber-Sher's case and corresponds to $\tan{\beta}=\infty$.

But for  $\lambda^2 > g_Z^2/2+g_{\eta}^2/3$, the maximum is obtained for:
\begin{equation}
\tan^2{\beta}=\frac{\lambda^2-\frac{g_Z^2}{2} +\frac{g_{\eta}^2}{12}}
{\lambda^2-\frac{g_Z^2}{2} -\frac{g_{\eta}^2}{3}}
\end{equation}
and a numerical evaluation of this expression requires a RGE approach
to fix a maximum value of $\lambda$.
This led to a bound on $M_{H_1}$ of qualitatively 170
 GeV at the time of the original Drees derivation in
which, we insist, the top mass was assumed to be of approximately 40 GeV and
the
hypothesis $\lambda^2/4\pi \leq 1$ at the scale $10^{16} $ GeV was used.

As one can see, the role of the  ratio $\lambda A/v_3$ is thus rather
 crucial in this game, since the difference between the two bounds is
( from an experimental point of view) indeed dramatic.
The aim of this paper is that of reconsidering the whole problem of
 the determination of a bound for $M_{H_1}$, in the most
general configuration in the parameter space for the
$\eta$ model, at the next one
loop order of perturbation theory, particularly taking into account the
fact that the top is now known to be heavier than
$\sim 90$ GeV from the last CDF limits [17].
For sake of comparison with the previous tree level
estimates, we shall still divide the parameter space
into two regions, corresponding to whether
the condition  $\lambda A / v_3 \ll 1$ is satisfied or not,
although in fact the "true" bound should always
be derived in the Drees' region.
\footnote{ Assuming $v_3\gg v_1,v_2 $ as from the previous discussion, the two
different regions can be classified from the
approximate expression, valid under this
assumption, of the ratio
$$
\frac{\lambda A}{v_3} = \frac{25}{18} g_{\eta}^2
\frac{\tan{\beta}}{1+ \tan{\beta}^2}
\frac{M_{ps}^2}{M_{z'}^2}
$$
showing  that the Drees' region does correspond to a substantially heavy
($M_{PS}\geq M_{Z'}$) pseudoscalar.}

In pratice, the expected modification of the bound at one
loop in the  Haber-Sher's region
is rather obvious if one believes that, in the
large $v_3$ limit , the exotic sector should simply
decouple from the conventional one.
In this case, the radiative corrections to the bound should simply come from
the top-stop sector.
The actual proof of this statement requires
a number of technicalities, that have been given in
a previous note [18].
The result is that, as expected, the one loop bound
in the Haber-Sher's region becomes:
\begin {equation}
M_{H_1}^2 \leq M_z^2 \left[\; 1+ \frac{16 g_{\eta}^2}{9 g_z^2}\; \right]+
\frac{3 \alpha}{2 \pi c_w^2 s_w^2} \frac{ M_{t}^4}{M_z^2}
\ln{ \frac{M_{\tilde {t}}^2}{M_t^2} }
\end{equation}
showing that the full one loop correction is just the large $\tan{\beta}$
limit of the corresponding correction in the case of the minimal  SUSY models
\footnote{ The fact that the bound remains finite i.e. the radiative correction
 is not of O($\alpha v_3$) is not occasional but is the consequence of a
general "screening" property valid
for a large class of SUSY models [19].}.

To evaluate the modification to the bound in the Drees'
 region, one has to start from the
expression of the modified eq.(7) at one loop.
This requires the explicit expression of the relevant
matrix elements at that order.
To perform the calculation, we have followed the effective
 potential approach [20] and we have first evaluated
  the contribution to
 the $ M^2$ matrix that is obtained by considering all the fermion and sfermion
 content of the model (including the eleven exotic states).
We have used the expressions of the masses of ref [21]; whenever this
 was possible, we have systematically neglected the various  D$-$terms
and/or terms of O($v/ v_3$).
Within these approximations, the starting expressions of the various masses
 become :

For the $top\!-\!stop$ system:
\begin{eqnarray}
M_t^2 & = &  h_t^2 v_2^2  \\
M_{\tilde {t}_R}^2 & = & m_{\tilde {t}_R}^2 + h_t^2 v_2^2 + \frac{5}{18}
g_{\eta}^2 v_3^2 +
h_t A v_2 + h_t v_1 v_3 \\
M_{\tilde {t}_L}^2 & = & m_{\tilde {t}_L}^2 + h_t^2 v_2^2 + \frac{5}{18}
g_{\eta}^2 v_3^2 -
h_t A v_2 - h_t v_1 v_3
\end{eqnarray}
with $h_t$ the top Yukawa coupling and $M_{\tilde {t}}$ the corresponding soft
mass
 Susy breaking (we shall assume, as it is usually done,
that $M_{\tilde{t}}=$O(TeV)) .
 Analogous expressions can be given for the $ b,\tilde {b}$ system,
 whose contribution
 turns out to be negligible at realistic $ \tan{\beta} $ values and will
 therefore be omitted in this first evaluation .

The quark exotic sector contributes with:
\begin{eqnarray}
M_h^2 & = & h_h^2 v_3^2 \\
M_{\tilde {h_{R}}}^2 & = & m_{\tilde {h}}^2 + h_h^2 v_3^2 - \frac{5}{36} k
g_{\eta}^2
 v_3^2 \\
M_{\tilde {h}_L}^2 & = & m_{\tilde {h}}^2 + h_h^2 v_3^2 - \frac{5}{9} k
g_{\eta}^2 v_3^2
\end{eqnarray}
where $h_{h}$ is the exotic Yukawa coupling which in principle can be large and
$m_{\tilde {h}}$ the soft mass;   in eq.(14),(15) k is a finite number that
 does not contribute appreciably to the result in any case.

The contribution of the other scalar mass sparticles for which the mixing is
 negligible
i.e.  $\tilde {f_{i}} = (\tilde {\upsilon},\; \tilde {\upsilon^{c}},\; \tilde
{e},
\; \tilde {e^{c}}$) is :
\begin{equation}
M_{\tilde {f_{i}}}^2 = m_{\tilde {f_{i}}}^2 + m_{f_{i}}^2 + \frac{5}{6}
g_{\eta}^2
Y_{1}^{i} v_3^2
\end{equation}
with $Y_{1}^{i}$ the extra U(1) hypercharges [22].

    Starting from eq. $ (10)\!-\!(16)$, inserting them in the effective
potential written
 as usually in the form [6]
\begin {equation}
  \Delta V(Q^2)= \frac{1}{64\pi^2} StrM^4  \ln [\, \frac{M^2}{Q^2} -
 \frac{3}{2} \,]
\end{equation}
   and reevaluating the minimum conditions , one arrives at the one
loop expressions
of the $ M^2$ matrix elements.

The results of our procedure are shown in the next formulae .
One sees that the formal $ v_3 $ dependence of the tree level expressions is
retained, and one can still write :
\begin {equation}
 a_{ij}^{(1)}  =
\alpha_{ij}^{(1)} v_3 + \beta_{ij}^{(1)} \;\;\; (i,j)\not=(3,3)
\end{equation}
\begin {equation}
a_{33}^{(1)}  =  \gamma_{33}^{(1)} v_3^2 + \delta_{33}^{(1)}
\end {equation}
where $ (\alpha,\beta,\gamma)_{ij} \;$ do not depend on $ v_3$ (and
$\; \delta_{33}
\sim 1 / v_3 $ ).
In particular, we find (the upper $ 1\!-$index denotes the complete quantity
 at one
loop; the same quantity without upper index is meant to be the tree level
expression, with renormalized couplings):
\begin{eqnarray}
\frac{\alpha_{11}^{(1)}}{\alpha_{11}} & = & \frac{\alpha_{12}^{(1)}}
{\alpha_{12}} =\frac{\alpha_{22}^{(1)}}{\alpha_{22}}  =  1- \frac{3}{16\pi^2}
 h_t^2 \ln{ \frac{M_{\tilde {t}}^2}{Q^2} } \\
\alpha_{13}^{(1)} & = & \alpha_{13}+ \frac{3}{8\pi^2} h_t^2 \lambda^2 v_1
\ln{ \frac{M_{\tilde {t}}^2}{Q^2} } \\
\alpha_{23}^{(1)} & = & \alpha_{23}+\frac{5}{24\pi^2} h_t^2 g_{\eta}^2 v_2
\ln{ \frac{M_{\tilde {t}}^2}{Q^2}} \\
\gamma_{33}^{(1)} & = & \gamma_{33}+ \frac{3}{4\pi^2} h_h^4 \ln{
\frac{M_{\tilde {h}}^2}{M_{h}^2}} + (D\!-\!terms>0)  \\
\frac{\beta_{11}^{(1)}}{ \beta_{11}} & = & \frac{\beta_{12}^{(1)}}{\beta_{12}}
  =  \frac{\delta_{33}^{(1)}}{\delta_{33}}  =  1 \\
\beta_{22}^{(1)} & = & \beta_{22}+ \frac{3}{4\pi^2} v_2^2 h_t^4
\ln{\frac{M_{\tilde {t}}^2}{M_t^2}} \\
\beta_{13}^{(1)} & = & \beta_{13}+ \frac{3}{16\pi^2} h_t^2 \lambda A v_2 \ln{
\frac{M_
{\tilde {t}}^2}{Q^2}} \\
\beta_{23}^{(1)} & = & \beta_{23}+ \frac{3}{16\pi^2} h_t^2 \lambda A v_1
\ln{\frac{
M_{\tilde {t}}^2}{Q^2} }
\end{eqnarray}
and in all the logarithms $M_{\tilde {t}_L}^2 = M_{\tilde {t}_R}^2 \equiv
M_{\tilde {t}}^2$ is
 used.
In eq.(23) we have called $ ``D\!-\!terms''$ the sum of all the contributions
of
this kind coming from the corresponding terms in the sfermion masses .
They will have little role in the numerical results, and we do not write
down their explicit form.

{}From the previous expression of the relevant matrix elements it would
be possible to compute numerically
the value of the light Higgs mass for any given choice of the parameters.
For what concerns the determination of a bound ,
though, the approach is relatively simpler since one
immediately realizes that the contribution to eq.(7) coming
from the (1,3) and (2,3) non diagonal elements is still
 definitely negative and as such it can be neglected for this specific purpose.
Since the exotic  contribution is essentially
contained in these terms, this means that
for the specific determination of the bound
it will still be possible to ignore it.
This remarkable simplification allows us to concentrate our analysis
on the reduced matrix
$M_{i,j}^2$, i , j = 1 , 2 , and to look
 for its maximum in the considered region.

To derive a maximum for the "reduced" component of eq.(7)
 we have proceeded as follows.
First, we have computed the modified relevant
 quantity at one loop, i.e. including the
(largely dominant) top-stop contribution,
 that coincides with the one appearing in eq.(9).
The modified one loop bound becomes therefore the following:
\begin {equation}
M_{H_1}^2 \leq \frac{ M_z^2}{(1+ \tan^2{\beta})^2}
           [ ( ( \tan^2{\beta}-1)^2 + \frac{g_{\eta}^2}{9 g_Z^2}
(4 \tan^2{\beta} + 1)^2 ) + \frac{8 \lambda^2}{g_Z^2} \tan^2{\beta} ]+
\frac{3 \alpha}{2 \pi c_w^2 s_w^2} \frac{ M_{t}^4}{M_z^2}
\ln{ \frac{M_{\tilde {t}}^2}{M_t^2} }
\end{equation}
The maximization of eq.(28) follows from the same
 assumptions that were used at tree level, with one crucial
difference coming from the extra positive
term that always increases with $M_t$
as $\sim M_t^4$.
On the contrary, the contribution to the maximum coming from the first two
terms on the r.h.s. of eq.(28) decreases with $M_t$ for not
too large $M_t$ values.
This is due to the fact that, in order to maximize $\lambda$, we shall
follow the convention of imposing "perturbative saturation" i.e.
$\lambda^2 / 4 \pi \leq 1$ at the $\Lambda=10^{16}$ GeV scale .
{}From the relevant RGE for $\lambda$ and $h_t$:
\begin {equation}
\frac{d \lambda}{d t} \simeq \frac{\lambda}{8 \pi^2}( -\frac{3}{2}g_L^2
-\frac{1}{2}
g_Y^2 -\frac{7}{6} g_{\eta}^2 + \frac{3}{2} h_t^2 +2 \lambda^2)
\end{equation}
\begin {equation}
\frac{d h_t}{d t} \simeq \frac{h_t}{8 \pi^2} (-\frac{8}{3} g_{strong}^2 -
\frac{3}{2} g_L^2 -\frac{13}{18} g_Y^2 -\frac{2}{3} g_{\eta}^2+
3 h_t^2 +\frac{1}{2} \lambda^2)
\end{equation}
where $t =\log{Q / \Lambda} $,
it follows then immediately that the value of $\lambda(Q)$
  decreases with $h_t$ in a way which
is shown in Fig.[1] \footnote{To derive this conclusion
 we have neglected in the RGE
the effect coming from the exotic coupling $h_E$, since the latter would
give in any case negative contributions.}.
Since $\tan{\beta}$ at the maximum is related to $\lambda$ from eq.(8), and
$h_t$ is related to $m_t$ and $\tan{\beta}$ by the equation:
\begin{equation}
 M_t^2= h_t^2 v^2 \frac{\tan^2{\beta}}{1+\tan^2{\beta}}
\end{equation}
this means that at the maximum all the various parameters i.e., $\lambda,
\tan{\beta}$ and $h_t$ will be expressed in terms of $M_t$.
In particular, the $\lambda$ dependence on $M_t$, that will crucial for
our conclusions, is shown in Fig.[2].

One can now proceed in the following way.
Of the three terms that appear in eq.(28), the third one
is clearly increasing with $M_t$.
The second O($\lambda^2$) terms decreases with $M_t$; this term would reproduce
essentially the Drees' bound at tree level, and is numerically
 dominant for relatively small $M_t$ values.
The first term increases with $M_t$, but remains relatively depressed
compared to the second one for small $M_t$ values.
This would exactly reproduce the Haber-Sher's bound for
 large $M_t$ values ( corresponding to
$\tan{\beta} \rightarrow \infty$)
when it becomes conversely dominant over the second one.

The overall picture is represented in Fig.[3], where we show
the mass bound as a function of the top mass in the full line, which is the
representation of eq.(28) with the constraint on $\tan \beta$ of eq.(8)
\footnote{It is interesting to note that the $M_t$ dependence of our bound  is
very
similar to the corresponding one in the "Minimal non Minimal" SUSY model
 recently examined in ref.[23].}.
The dotted line represents the upper bound of eq.(28) without the explicit
contribution of the radiative correction $\propto M_t^4$ and shows the
importance of the latter expecially for large values of $M_t$.
Finally, the point-dotted line represents the radiative corrected upper
bound in the Haber-Sher's region, eq.(9), showing that for large $M_t$ values
 the bounds, in the two regions, merge into one identical result.

One seees that for $M_t\leq 2 M_W$ GeV there exists
a bound for the model that would correspond
to the modification of the Drees' bound and decreases with
$M_t$, remaining always smaller than $\sim(160$GeV) for $M_t\geq 90$ GeV.
This would be in qualitative agreement with a general statement recently
made by Kane et al. [24].
For larger $M_t$ values, the bound begins to
 increase with $M_t$.
Assuming the limit on $M_t$ derivable from the last LEP1 data
 $M_t \leq 200$ GeV, we obtain for this model the corresponding perturbatively
saturated bound
\begin{equation}
M_{H_1}\leq 160 \;\; GeV
\end{equation}
Thus, if the next LEP2 analyses failed to find a light scalar and set
 a qualitative limit $M_{H_1}\geq M_Z$, the
model would predict a light scalar in a region that might be
thoroughly ( and particularly
 well )  investigated by future higher energy $e^+ e^-$
linear colliders [25].

Acknowledgements

We have greatly benefitted from several discussion with  M.Drees.

\vfill
\newpage

REFERENCES

[1] See e.g. for a discussion of this point L.E.Ib\'{a}\~{n}ez ,
$CERN-TH.5982/91 $

[2] E.Cremmer, S.Ferrara, L.Girardello, A.Van Proeyen ,
 Nucl.Phys. B 212 (1983) 413

L.E.Ib\'{a}\~{n}ez, Phys.Lett. 118 B 1982 73 ; Nucl.Phys. B 218 (1983) 514

R.Barbieri, S.Ferrara, C.Savoy, Phys.Lett. 119 B (1982) 343

[3] See e.g. L.Hall, L.Randall, Phys.Rev. Lett. 65 (1990) 2939

[4] L.E.Ib\'{a}\~{n}ez, Nucl.Phys. B 218 (1983) 514

L.E.Ib\'{a}\~{n}ez, C.L\'{o}pez, Phys.Lett. 126 B (1983) 54

L.Alvarez Gaum\'{e}, J.Polchinsky , M.Wise, Nucl.Phys. B 221 (1983) 495

 J.Ellis, J. Hagelin, D.Nanopoulos, K.Tanvakis, Phys.Lett. 125 B (1983) 275

[5] H.E.Haber, R.Hempfling, Phys.Rev. Lett.66 (1991) 1815

J.Ellis, G.Ridolfi, F.Zwirner,  Phys.Lett.  B 257 (1991) 83

R.Barbieri, M.Frigeni, F.Caravaglios, Phys.Lett. 258 B (1991) 395

[6] S.Coleman, E.Weinberg, Phys.Rev. D 7 (1973) 1888

[7] See for instance the discussion of P.Janot, Proceedings of the $27^{th}$
Rencontre de Moriond:

"Electroweak interactions and unified theories", pag.68 ,J.Tran Thanh Van
editor

[8] M.Drees,M.M.Nojiri, Phys.Rev. D 45 (1992) 2482

[9] H.E.Haber, M.Sher, Phys. Rev. D 35 (1987) 2206

[10] Y. Hosatani, Phys.Lett. 129 B (1983) 193

[11] V.Barger, K.Whisnant, Int.J.of Mod.Phys. A, Vol. 3, No.8 (1988) 1907

J.L.Hewett, T.Rizzo, Phys.Rep. 183 (1989) 194

[12] M.Drees, N.K.Falck, M.Gluck, Phys.Lett. 167 B (1986) 187

M.Drees, Phys. Rev. D 35 (1987) 2910

[13] See  K.Maeshima ,Proceedings of the $27^{th}$ Rencontre de Moriond:

"Electroweak interactions and unified theories", pag.125 ,J.Tran Thanh Van
editor

[14] J.Layssac, F.M.Renard, C.Verz,egnassi, Phys.Lett.B 287 (1992)  267

[15]  J.Nash, Proceedings of the $27^{th}$ Rencontre de Moriond:

"Electroweak interactions and unified theories", pag.9 ,J.Tran Thanh Van editor

[16] A.Djouadi, A.Leike, T.Riemann, D.Schaile, C.Verzegnassi, Z Phys.C 56
(1992) 289

[17] CDF collab. , F.Abe et al., Fermilab-Pub 90-1373 (1990)

[18] D.Comelli, C.Verzegnassi, preprint DESY 92-089

[19] D.Comelli, C.Verzegnassi,preprint DESY 92-109 to appear on Phys. Rev.D

[20] G.Gamberini, G.Ridolfi, F.Zwirner, Nucl.Phys. B 331 (1990) 331

[21] J.Ellis, K.Enqvist, D.V.Nanopoulos, F.Zwirner, Nucl.Phys. B 276 (1986) 14

[22] E.Ma, Phys.Rev. D 36 (1987) 274

[23] J.R.Espinosa, M. Quiros, Phys.Lett. B266 (1992) 92

D.Comelli, preprint HEP-PH/9211230

J.R.Espinosa, M. Quiros, preprint HEPPH/9212305

[24] G.L.Kane, C.Kolda, J.D.Wells, Preprint Univ.of Michigan  UM-TH-92-24

[25] See for a discussion the proceeding of the workshop:" $e^+ e^- $ Collision
at 500 GeV", DESY 92-123 A, P.M. Zerwas editor

\vfill
\newpage

FIGURE CAPTIONS

Fig.1

Values of $\lambda(M_Z)$ for variable $h_t(M_Z)$ calculated with the
RGE of eq.(29),(30) with the saturation condition $\lambda^2(\Lambda)/4 \pi
=1$ and $\Lambda=10^{16}$ GeV.

Fig.2

Dependence of  $M_t$ on $\lambda(M_Z)$ calculated after imposing the
condition of eq.(9) on $\tan \beta$ and the relation between $\lambda$
and $h_t$ fixed by by the RGE shown in Fig.[1].

Fig.3

Values of the light Higgs mass bound at variable $M_t$ for
$M_{\tilde {t}}=1$ TeV.
The full upper line represents the
bound of eq.(28) with the constraint of
eq.(8).
The dotted line represents the same equation
without the last contribution $\propto M_t^4$.
The point-dotted line shows the bound in the Haber-Sher's region, eq.(9).

\end{document}